\documentclass[final,5p,times,twocolumn]{article}

\usepackage[utf8]{inputenc}
\usepackage{natbib}
\usepackage{graphicx}
\usepackage[margin=0.5in]{geometry}

\usepackage{lineno,hyperref}
\modulolinenumbers[5]

\usepackage{epstopdf}
\usepackage{verbatim}
\usepackage{mathrsfs}
\usepackage{array}

\newcolumntype{H}{>{\setbox0=\hbox\bgroup}c<{\egroup}@{}}

\begin{document}

\title{An empirical consistent redshift bias: A possible direct reproducible observation of Zwicky's TL theory}

\date{}

\author{Lior Shamir\footnote{lshamir@mtu.edu}  \\ Kansas State University \\ 1701 Platt St \\ Manhattan, KS 66506, USA}

\maketitle

\abstract{
Recent advancements have shown tensions between observations and our current understanding of the Universe. Such observations may include the $H_o$ tension and massive galaxies at high redshifts that are older than what traditional galaxy formation models predicted. Since these observations are based on the redshift as the primary distance indicator, a bias in the redshift may explain these tensions. While the redshift follows an established model, when applied to astronomy it is based on the assumption that the rotational velocity of the Milky Way galaxy relative to the observed galaxies has a negligible effect on the redshift. But given the mysterious nature of the physics of galaxy rotation, that assumption should be tested. The test is done by comparing the redshift of galaxies rotating in the same direction relative to the Milky Way to the redshift of galaxies rotating in the opposite direction relative to the Milky Way. The results show that the mean redshift of galaxies that rotate in the same direction relative to the Milky Way is higher than the mean redshift of galaxies that rotate in the opposite direction. Additionally, the redshift difference becomes larger as the redshift gets higher. The consistency of the analysis was verified by comparing data collected by three different telescopes, annotated using four different methods, released by three different research teams, and cover both the Northern and Southern ends of the galactic pole. All datasets are in excellent agreement with each other, showing consistency in the observed redshift bias. Given the ``reproducibility crisis'' in science, all datasets used in this study are publicly available, and the results can be easily reproduced. The observation could be a first direct empirical reproducible observation for the Zwicky's ``tired-light'' model.
}



\section{Introduction}
\label{introduction}

he unprecedented imaging power of JWST has revealed new information about the Universe that is not aligned with some of the current fundamental cosmological assumptions. For instance, the presence of galaxies at redshifts higher than 13 \citep{curtis2023spectroscopic} or even as high as 15 \citep{whitler2023ages} was not expected according to previous assumptions \citep{cowley2018predictions}. Mature massive galaxies observed at redshift of $\sim$11 \citep{glazebrook2024massive} also challenge the cosmological model and the history of the Universe. The existence of massive galaxies at high redshifts was reported also before JWST was launched \citep{neeleman2020cold}, and JWST provided more and deeper instances of such galaxies, showing that mature galaxies were prevalent in the early Universe. Another example of a puzzling unexplained observation is the $H_0$ tension, reflected by two different messengers that provide two different expansion rates and ages of the Universe \citep{wu2017sample,mortsell2018does,bolejko2018emerging,davis2019can,pandey2020model,camarena2020local,di2021realm,riess2022comprehensive}. Since both measure the same Universe, it can be assumed that one of these measurements is biased.  
 
These puzzling observations introduce a challenge to cosmology: If the distance indicators are fully accurate, the standard cosmological theories are incomplete. If the current standard cosmological theories are complete, then the distance indicators might not be fully accurate. That is, either the standard cosmological theories need to be revised, or the redshift as a distance indicator needs to be revised, but the two might not be able to co-exist without modifications. While some solutions include alternative cosmological models, it has also been suggested that the tensions can be solved by changing the redshift model \citep{crawford1999curvature,pletchermature,gupta2023jwst,lee2023cosmological,lovyagin2022cosmological}. 

The redshift of an astronomical object is expected to correspond to the linear velocity of the object relative to Earth. But in addition to the linear velocity, the measurement of the redshift can also be affected by the rotational velocity of the observed galaxy, or the rotational velocity of the Milky Way. The rotational velocity of a luminous object can affect the Doppler shift effect \citep{marrucci2013spinning,lavery2014observation,liu2019experimental}, but since the rotational velocity of a galaxy is far smaller than its linear velocity, it is assumed that the effect of the rotational velocity is negligible. For that reason the rotational velocity of galaxies is ignored in common redshift models. However, it should be reminded that the physics of galaxy rotation, and namely its rotational velocity, is still one of the most mysterious and most provocative phenomena in nature. 

While the puzzling nature of the physics of galaxy rotation was noted in the first half of the 20th century \citep{oort1940some}, only five decades later it became part of the mainstream science \citep{rubin1983rotation,el2020aedge}. Common explanations include dark matter \citep{rubin1983rotation} or modified Newtonian dynamics \citep{milgrom1983modification}, but numerous other explanations have also been proposed \citep{sanders1990mass,capozziello2012dark,chadwick2013gravitational,farnes2018unifying,rivera2020alternative,nagao2020galactic,blake2021relativistic,gomel2021effects,skordis2021new,larin2022towards}.  But despite over a century of scientific research, there is still no proven explanation to the nature of galaxy rotation \citep{sanders1990mass,mannheim2006alternatives,kroupa2012dark,kroupa2012failures,kroupa2015galaxies,arun2017dark,akerib2017,bertone2018new,aprile2018,skordis2019gravitational,sivaram2020mond,hofmeister2020debate,byrd2021spiral,haslbauer2022high,haslbauer2022has}.

The purpose of this experiment is to examine the impact of the galaxy rotational velocity on the redshift. That can be done by observing galaxies located around the Galactic pole, and comparing the redshift of galaxies that rotate in the same direction relative to the Milky Way to the redshift of galaxies that rotate in the opposite direction relative to the Milky Way. An observed difference in the redshift can indicate that the redshift is affected by the rotational velocity, and therefore the way it is implemented in astronomy is incomplete.

\section{A possible link between the galaxy rotational velocity and the redshift}
\label{link}

A preliminary possible link between the redshift and the rotational velocity of galaxies have been proposed through empirical observations showing that galaxies that rotate in the same direction relative to the Milky Way have different redshift compared to galaxies that rotate in the opposite direction relative to the Milky Way  \citep{universe10030129}. The analysis was based on several different datasets of galaxies acquired by two different telescopes, and annotated by their direction of rotation using several different annotation methods. Since galaxies with leading arms are extremely rare \citep{buta2003ringed}, the direction of rotation was determined by the arms of the galaxies. The analysis was based on galaxies at the $10^o\times10^o$ field centered at the Galactic pole, as well as the $20^o\times20^o$ field to increase the number of galaxies and consequently the statistical significance. The analysis was done for both the Northern and Southern ends of the Galactic pole \citep{universe10030129}.

The catalogs used in the experiment included galaxies annotated by the {\it Ganalyzer} algorithm \citep{shamir2011ganalyzer,ganalyzer_ascl}, by the {\it SpArcFiRe} method \citep{Davis_2014,sparcfire_ascl}, and by the manual annotation of {\it Galaxy Zoo} such that the ``superclean'' criteria of 95\% agreement between the annotations \citep{lintott2008galaxy} was used. All of these methods determine the spin direction of the galaxies based on analysis if the shape of the galaxy arms. The datasets are all available publicly. SDSS and DESI galaxies annotated by the {\it Ganalyzer} algorithm are available at \url{https://people.cs.ksu.edu/~lshamir/data/zdifference/}. Galaxies annotated by {\it SpArcFiRe} are available at \url{https://people.cs.ksu.edu/~lshamir/data/sparcfire/}. The annotations of {\it Galaxy Zoo} can be accessed through SDSS CAS server at \url{http://casjobs.sdss.org/CasJobs/default.aspx}. Table~\ref{main_dataset} shows the mean redshift of galaxies that rotate in the same direction and in the opposite direction relative to the Milky Way.

\begin{table*}[h]
\caption{The mean redshift of galaxies that rotate in the same direction relative to the Milky Way galaxy  (MW) and the mean redshift of galaxies that rotate in the opposite direction relative to the Milky Way (OWM). The $p$ values are the one-tailed $p$ values determined by the Student t-test. All datasets are publicly available. The experiments are described in detail in \citep{universe10030129}.}   
\label{main_dataset}
\centering
\scriptsize
\begin{tabular}{lccccccccc}
\hline
Survey & Pole               & Field             & Annotation  & \# MW  & \# OMW  & $Z_{mw}$  & $Z_{omw}$  & $\Delta$z   & t-test \\
           &     & size  ($^o$)  &                  &             &               &                  &                    &                 &    p     \\ 
\hline
SDSS & North & 10$\times$10     &   Ganalyzer    & 204   &  202  & 0.0996$\pm$0.0036    &  0.08774$\pm$0.0036     & 0.01185$\pm$0.005 & 0.01 \\  
SDSS & North & 20$\times$20     &   Ganalyzer    &  817   &  825  & 0.09545$\pm$0.0017  &  0.08895$\pm$0.0016     &  0.0065$\pm$0.0023 & 0.0029 \\ 
SDSS & North & 20$\times$20     &   Galaxy Zoo  & 154   & 135   & 0.07384$\pm$0.004    &  0.06829$\pm$0.0035    & 0.0056$\pm$0.0053   & 0.15 \\
SDSS  & North & 10$\times$10     &  {SpArcFiRe}             &  710       &  732  &  0.07197$\pm$0.0015  &  0.06234$\pm$0.0014 & 0.00963$\pm$0.002 & $<$0.0001 \\  
SDSS & North & 10$\times$10      & {SpArcFiRe}   &   728       &   709  &  0.06375$\pm$0.0014  & 0.07191$\pm$0.0014  & -0.00816$\pm$0.002 & $<$0.0001 \\   
           &       &                             &    Mirrored      &             &               &                              &                                   &                 &         \\ 
SDSS  & North &  20$\times$20     &  {SpArcFiRe}   &   2903   &  2976  & 0.07285$\pm$0.0007  &  0.07116$\pm$0.0007 &  0.00169$\pm$0.001 & 0.04 \\  
SDSS  & North &   20$\times$20   & {SpArcFiRe}  &   3003   &   2914  & 0.07113$\pm$0.0007  &  0.07271$\pm$0.0007  &  -0.00158$\pm$0.001 & 0.05 \\  
           &       &                             &    Mirrored      &             &               &                              &                                   &                 &         \\ 
DESI & South & 10$\times$10    &   Ganalyzer    &    414   &  376   &  0.1352$\pm$0.0027 & 0.1270$\pm$0.0025    & 0.0082$\pm$0.0036 & 0.018 \\   
DESI & South & 20$\times$20    &   Ganalyzer    &    1702 &  1681  &  0.1317$\pm$0.0013 & 0.1273$\pm$0.0014   & 0.0044$\pm$0.0018 & 0.008 \\  

\hline
\end{tabular}
\end{table*}

As the table shows, all catalogs show lower redshift for galaxies that rotate in opposite direction relative to the Milky Way. With the exception of the small Galaxy Zoo catalog, in all cases the difference is statistically significant. The observed $\Delta$z difference was higher when using the smaller $10^o\times10^o$ field centered at the Galactic pole compared to the larger $20^o\times20^o$ field. That can be explained by the difference in the relative rotational velocity that is expected to increase when the observed galaxies are closer to the Galactic pole. The {\it SpArcFiRe} algorithm was used with both the original images and with the mirrored images. That was done due to the reported bias of the {\it SpArcFiRe} method \citep{hayes2017nature}. {\it SpArcFiRe} provided a higher number of galaxies, but on the expense of the accuracy of the annotation, leading to a smaller absolute $\Delta$z though stronger statistical signal due to the higher number of galaxies \citep{universe10030129}. The {\it Ganalyzer} algorithm has a simple ``mechanical'' symmetric nature \citep{shamir2011ganalyzer,shamir2021particles,shamir2022analysis}, and therefore mirroring the galaxies did not change the results \citep{universe10030129}. When using {\it SpArcFiRe} the results changed when the galaxy images were mirrored, but the change was not substantial, except for the expected inverse $\Delta$z observed when the galaxy images were mirrored.

The p-values shown in Table~\ref{main_dataset} are based on Student t-test, which assumes normal distribution of the redshift. Because the redshift distribution does not necessarily follow normal distribution, the Student t-test might not provide the real probability to have such distribution by chance. To verify the statistical significance, a simulation analysis was used such that the galaxies were separated randomly into two groups, and the difference between the mean redshift of each of the two groups of galaxies was computed. That was repeated 100,000 times with the dataset of 1,642 galaxies of the second row in Table~\ref{main_dataset}. In 307 of the runs the difference between the first group of galaxies and the second group was larger than 0.0016, providing a probability of 0.0031 to occur by chance \citep{universe10030129}.

The DESI Legacy Survey data was collected around the Southern galactic pole, and therefore galaxies in that field that rotate in the same direction relative to the Milky Way would seem to rotate in the opposite direction compared to galaxies in the Northern galactic pole that rotate in the same direction relative to the Milky Way. That provides an additional verification that the difference is not a feature of some unknown behavior of the galaxy annotation methods, as such effect should have been consistent across the sky, rather than flip and provide inverse results in the two ends of the Galactic pole \citep{universe10030129}. An additional control experiment was performed by using SDSS galaxies from the same source that rotate in opposite directions in fields that are perpendicular to the Galactic pole. For instance, in the 20$\times$20 degree fields centered at $(\alpha=102^o,\delta=0^o)$ the $\Delta$z was 0.00022$\pm$0.004, which is statistically insignificant. 


Table~\ref{spec_dif} shows the differences between the flux in different wavelengths. The flux differences between galaxies that rotate in opposite directions show a difference of $\sim$10\% in the different filters, and the absolute difference is larger when the wavelength is shorter \citep{universe10030129}.

\begin{table}
\caption{Flux differences in different filters between galaxies that rotate in the same and in the opposite direction relative to the Milky Way. The data are the galaxies and the field shown in the first row of Table~\ref{main_dataset}. The $p$ values are the two-tailed Student t-test $p$ value.}   
\label{spec_dif}
\centering
\scriptsize
\begin{tabular}{lcccccc}
\hline
Band            & MW           & OMW  & $\Delta$   & t-test P \\
\hline
spectroFlux\_g  &   25.969$\pm$0.8669        &  28.554$\pm$1.0918     & -2.585  &  0.063 \\  
spectroFlux\_r   &   53.2433$\pm$1.765        & 58.6214$\pm$2.3422    &  -5.378  &  0.066 \\  
spectroFlux\_i   &   77.4189$\pm$2.513        &  85.0868$\pm$3.407     &  -7.667  &  0.067 \\  

\hline
\end{tabular}
\end{table}

\section{Analysis of HSC DR3 galaxies}
\label{hsc}

The analysis shown in Section~\ref{link} and in \citep{universe10030129} shows a consistent but relatively small redshift bias. But these analyses are based on galaxies imaged mostly by SDSS, with one experiment with galaxies obtained through DESI Legacy Survey. Therefore, most galaxies used for the experiments included in Table~\ref{main_dataset} are of relatively low redshift. That makes it difficult to profile the change in $\Delta$z when the redshift changes. That is, if $\Delta$z increases at higher redshift, it can provide an indication that the redshift bias is higher for high-redshift galaxies.

To profile the dependence between the $\Delta$z and the redshift, an experiment was performed with the third data release (DR3) of the Hyper Suprime-Cam (HSC). Using the 8.2m Subaru telescope, HSC provides high details of galaxies with higher redshifts compared to galaxies imaged by SDSS or DESI Legacy Surveys. That allows to annotate galaxies with higher redshift by their direction of rotation, and test whether the redshift bias increases with the redshift.

The galaxies used in the experiment include all HSC DR3 galaxies that have redshifts in SDSS DR17. That included 101,415 galaxies with redshift of $z<0.3$. The purpose of the redshift limit was to avoid galaxies that cannot be annotated accurately by their visible spin patterns. The galaxies were annotated by the {\it Ganalyzer} algorithm \citep{shamir2011ganalyzer,ganalyzer_ascl} as described in \citep{universe10030129}. That led to a clean dataset of 13,477 galaxies. 

\begin{figure}[h]
\centering
\includegraphics[scale=0.60]{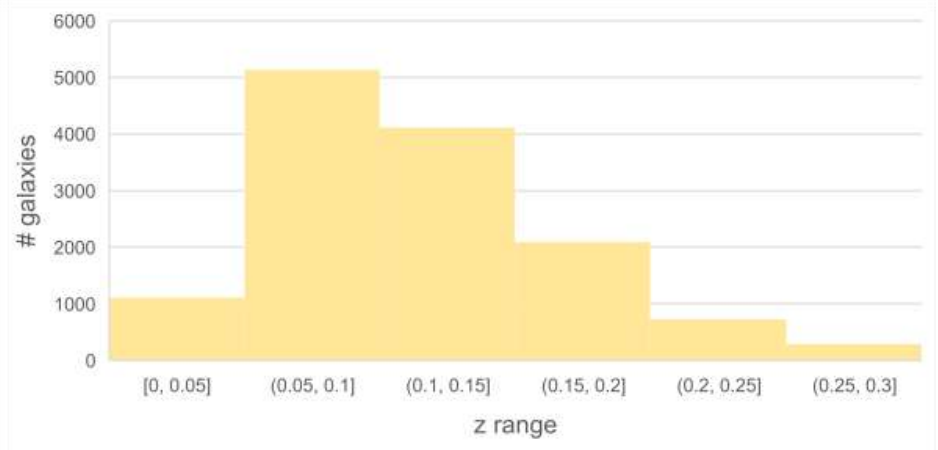}
\caption{Distribution of the reshift of the HSC DR3 galaxies.}
\label{redshift}
\end{figure}

The smaller dataset of annotated galaxies compared to the initial dataset is expected since not all galaxies are spiral, and not all spiral galaxies have an identifiable direction of rotation. Manual inspection of randomly selected 100 galaxies showed that all galaxies are annotated accurately. The dataset can be accessed at \url{https://people.cs.ksu.edu/~lshamir/data/zdifference_hsc/}. The redshift distribution of the galaxies is displayed in Figure~\ref{redshift}. Although most galaxies have a relatively low redshift, the catalog still contains more than 1000 galaxies with $z>0.2$. 

\begin{figure}[h]
\centering
\includegraphics[scale=0.60]{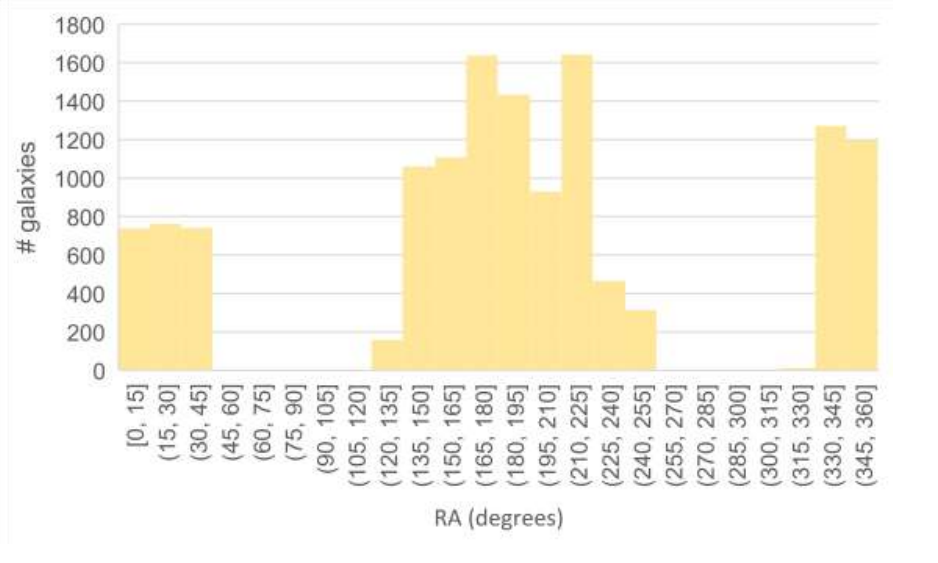}
\caption{Distribution of the RA in the dataset of HSC DR3 galaxies.}
\label{ra}
\end{figure}

While providing better image quality and depth compared to the catalogs summarized in Section~\ref{link}, the downside of the HSC catalog used here is that it does not cover neither the South nor North end of the Galactic pole. Therefore, the galaxies used for the analysis are galaxies that are closer to the South end of the Galactic pole, as well as galaxies that are close to the North end of the Galactic pole. The declination of the galaxies range between $-6.58^o$ to $53.18^o$. The distribution of the right ascension is shown in Figure~\ref{ra}. The catalog contains 4,724 galaxies that are within 60$^o$ from the Southern Galactic pole, and 8,753 galaxies that are within 60$^o$ from the Northern Galactic pole. 

Tables~\ref{south_pole} and~\ref{north_pole} show the mean redshift of galaxies that rotate in the same direction relative to the Milky Way and in the opposite direction relative to the Milky Way in the Southern and Northern Galactic poles, respectively. The tables show the difference between the mean z among all galaxies, but also  the differences when the galaxy population is separated into 0.1 redshift ranges. As the table shows, in both ends of the Galactic pole there is a statistically significant difference between the redfshift of galaxies that rotate in the same direction as the Galactic pole, and galaxies that rotate in the opposite direction relative to the Galactic pole.
  
The direction of rotation of the galaxies observed in the opposite sides of the Galactic pole are inverse, such that a galaxy located around the North end of the Galactic pole and rotates in the same direction relative to the Milky Way would seem to rotate in the opposite direction if it was located in the South end of the Galactic pole. That is naturally taken into account in the analysis, but the consistency between both ends of the Galactic pole shows that the differences are not driven by an unknown bias in the annotation, as such bias should have been consistent in both ends of the Galactic pole, and not expected to flip based on the locations of the galaxies in the sky.
 
As discussed in Section~\ref{link}, the p-values are based on Student t-test, and therefore on the the assumption that the redshifts are distributed normally. Since that assumption might not necessarily be correct, a simulation was performed such that the galaxies were separated into two groups randomly, and the mean redshift of one group was compared to the mean redshift of the galaxies of the other group \citep{universe10030129}. After 100,000 tests, the galaxies close to the Southern Galactic pole were separated into two random groups with mean redshift difference larger than 0.00363 in 518 runs. The same experiment when using the galaxies closer to the Northern Galactic pole showed a mean redshift difference larger than 0.002451 in 230 of the runs. These simulation experiments show that the probability to have such separation by mere chance is far below 1\%.
 
\begin{table*}[h]
\caption{Redshift of the galaxies around the Southern galactic pole that rotate in the same direction relative to the Milky Way and in the opposite direction relative to the Milky Way. The p values are the Student t-test probabilities to have such difference are stronger by chance.}  
\label{south_pole}
\centering
\scriptsize
\begin{tabular}{lcccccc}
\hline
z range & \# MW & \# OMW & $Z_{mw}$  & $Z_{omw}$  & $\Delta$z   & t-test P \\
\hline
0-0.1     & 871     &  917      &     0.072788$\pm$0.0006  & 0.071198$\pm$0.0006    & 0.001589$\pm$0.0009  &  0.03 \\
0.1-0.2  & 1,100   & 1,144    &     0.149292$\pm$0.0008  & 0.145834$\pm$0.0008  & 0.003458$\pm$0.001   & 0.001 \\
0.2-0.3  & 342     &  350       &    0.242174$\pm$0.0015  & 0.236287$\pm$0.001      & 0.005886$\pm$0.002   & 0.0006 \\
All         & 2,313  & 2,411     &    0.13421$\pm$0.001      & 0.13058$\pm$0.001       &  0.00363$\pm$0.0014  & 0.005  \\
\hline
\end{tabular}
\end{table*}

\begin{table*}[h]
\caption{Redshift of the galaxies around the Northern galactic pole that rotate in the same direction relative to the Milky Way and in the opposite direction relative to the Milky Way.}  
\label{north_pole}
\centering
\scriptsize
\begin{tabular}{lcccccc}
\hline
z range & \# MW & \# OMW & $Z_{mw}$  & $Z_{omw}$  & $\Delta$z   & t-test P \\
\hline
0-0.1     & 2,202     &  2,255      &     0.070232$\pm$0.0004  & 0.069493$\pm$0.0004    & 0.000739$\pm$0.0006  &  0.095 \\
0.1-0.2  & 1,949     & 2,021      &      0.138494$\pm$0.0006  & 0.134705$\pm$0.0005    & 0.003789$\pm$0.0008  & 0.00005 \\
0.2-0.3  & 166       &  160        &      0.228586$\pm$0.0018  & 0.221593$\pm$0.0014     & 0.006993$\pm$0.0023  & 0.0012 \\
All         & 4,317    & 4,436      &      0.10714$\pm$0.0006    & 0.104689$\pm$0.0006     & 0.002451$\pm$0.0008  & 0.0019  \\
\hline
\end{tabular}
\end{table*}

\begin{figure}
\centering
\includegraphics[scale=0.75]{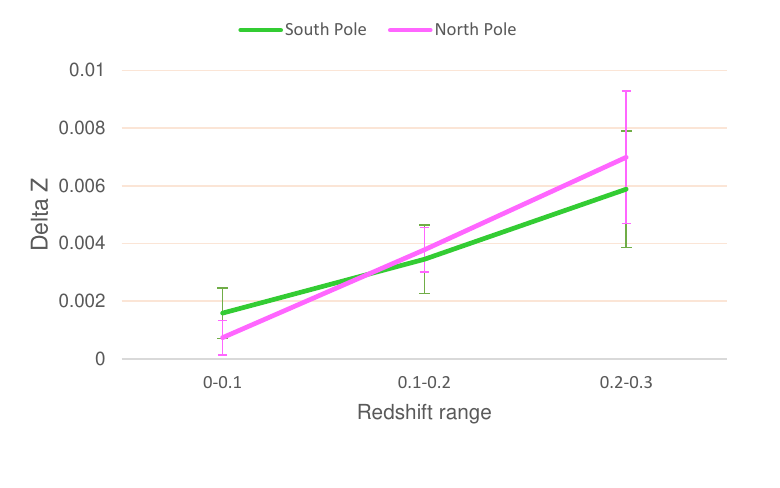}
\caption{$\Delta$z in different redshift ranges in the South and North end of the Galactic pole. Both ends of the Galactic pole show a very similar $\Delta$z increase.}
\label{increase}
\end{figure}

The simple separation of the analysis into several redshift ranges shows that $\Delta$z increases as the redshift gets higher. That observation is consistent in both ends of the Galactic pole. Figure~\ref{increase} shows the $\Delta$z in the different redshift ranges. 

The statistical significance of the increase in $\Delta$z can also be determined by using a simple Pearson correlation after assigning the galaxies that rotate in the same direction relative to the the Milky Way with the value 1, and galaxies that rotate in the opposite direction relative to the Milky Way as -1.  When using galaxies closer to the Northern Galactic pole the Pearson correlation is 0.0296, and the probability to have such correlation by chance is $p$=0.02. When using galaxies closer to the Southern Galactic pole, the Pearson correlation is 0.02669, and the $p$ is 0.0064.

\section{Verification using a third-party dataset}
\label{third_party}

\begin{figure}[h]
\centering
\includegraphics[scale=0.60]{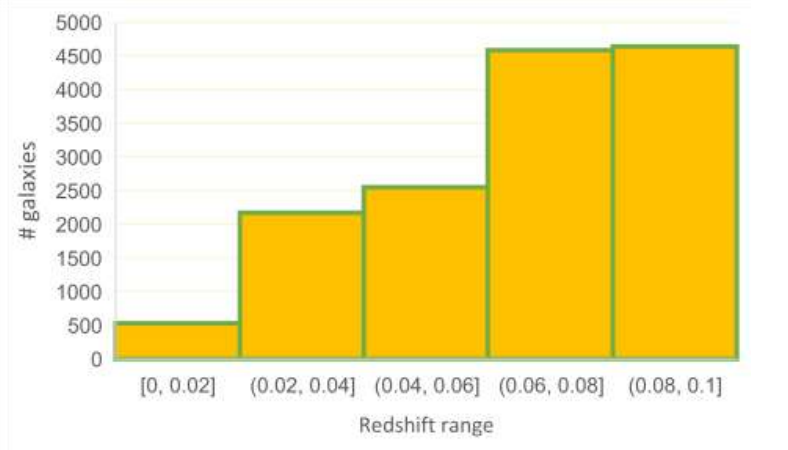}
\caption{Distribution of the reshift of the galaxies in the catalog of \citep{longo2011detection}. The redshift of all galaxies is smaller than 0.1.}
\label{longo_redshift}
\end{figure}

\begin{figure}[h]
\centering
\includegraphics[scale=0.60]{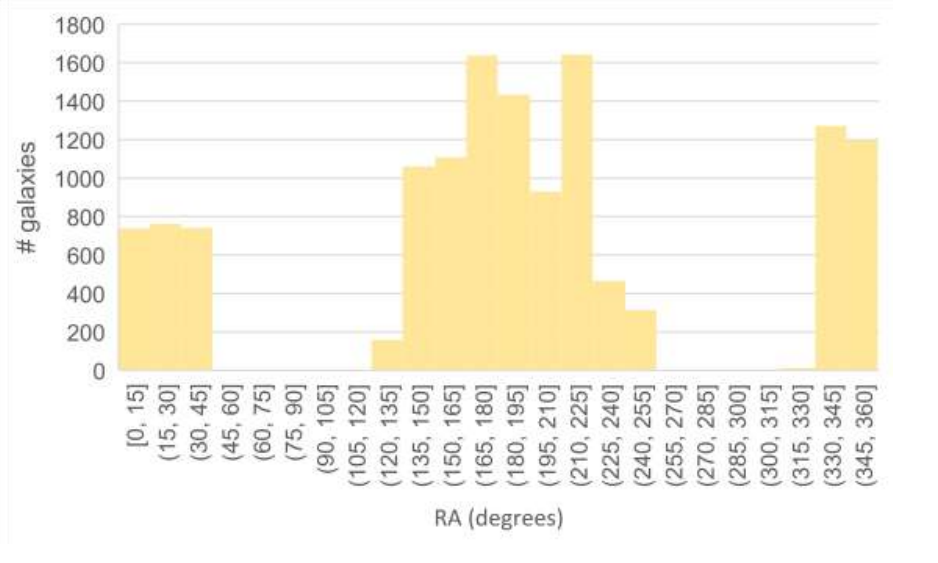}
\caption{Distribution of the RA in the catalog of \citep{longo2011detection}. Most galaxies are closer to the North end of the Galactic pole.}
\label{ra_longo}
\end{figure}

To further test the consistency of the bias, the observation described in Sections~\ref{link} and~\ref{hsc} was compared to a third party catalog \citep{longo2011detection}. That catalog was prepared by five undergraduate students who manually annotated SDSS galaxies, and the galaxy images were also mirrored to offset for a possible human bias \citep{longo2011detection}. While the purpose of that dataset was not to compare redshifts or test the impact of the rotational velocity of the Milky Way, 14,462 of the galaxies in that dataset has redshift, and therefore can be used for the analysis. The dataset can be accessed at \url{https://ars.els-cdn.com/content/image/1-s2.0-S0370269311003947-mmc1.txt}. The redshift of all galaxies that have spectra is smaller than 0.1, and the redshift distribution is displayed by Figure~\ref{longo_redshift}. The distribution of the RA is displayed in Figure~\ref{ra_longo}. As the figure shows, the majority of the galaxies are in relatively close proximity to the Northern Galactic pole.

A very simple analysis of the data shows that the mean redshift of the galaxies that rotate clockwise in that catalog is 0.06586$\pm$0.00029, while the mean redshift of galaxies rotating counterclockwise is 0.065$\pm$0.00029. The probability to have such difference by chance is 0.018. While the statistically significant redshift difference might seem unexpected, most of the galaxies are located near the Northern Galactic pole, and therefore the lower redshift for galaxies that rotate counterclockwise is in agreement with the results shown in Table~\ref{north_pole} using the HSC data taken from the Northern galactic pole, as well as the other datasets show in Table~\ref{main_dataset}. Table~\ref{north_longo} shows the mean z of galaxies that rotate in the same direction relative to the Milky Way in the catalog \citep{longo2011detection} and the mean z of galaxies that rotate in the opposite direction relative to the Milky Way in the Southern and Northern Galactic poles. 

\begin{table*}
\caption{Redshift of the galaxies annotated by \citep{longo2011detection} around the Southern and Northern galactic poles that rotate in the same direction relative to the Milky Way and in the opposite direction relative to the Milky Way. The older third-party dataset is used to confirm the results observed with the other datasets.}  
\label{north_longo}
\centering
\scriptsize
\begin{tabular}{lcccccc}
\hline
Hemisphere & \# MW & \# OMW & $Z_{mw}$  & $Z_{omw}$  & $\Delta$z   & t-test P \\
 & & &  &  &  & values \\
\hline
North    &   6450  & 6573 & 0.06598$\pm$0.0003 &  0.06498$\pm$0.0003 & 0.001$\pm$0.0004 & 0.01 \\
South    &  722    &  717  & 0.06516$\pm$0.0008  & 0.06469$\pm$0.0009 & 0.00047$\pm$0.0012 & 0.34  \\
\hline
\end{tabular}
\end{table*}

As the table shows, the $\Delta$z around the Northern Galactic pole agrees well with the $\Delta$z observed by HSC around the Northern Galactic pole at the same redshift range of 0-0.1, as shown in Table~\ref{north_pole}. The opposite end of the Galactic pole also shows that galaxies that rotate in the opposite direction relative to the Milky Way have lower redshift compared to galaxies that rotate in the opposite direction, but in the case of the \citep{longo2011detection} catalog the difference is not statistically significant, possible due to the low number of galaxies in that part of the sky. 

Since the redshift of the galaxies in that catalog is limited to $z<0.1$, a direct comparison to the redshift ranges used in Table~\ref{north_pole} is not possible. However, when using the galaxies around the Northern Galactic pole, the Pearson correlation coefficient between the direction of rotation and the redshift is 0.02033, and the probability to have this coefficient by chance is $\sim$0.01. That provides an indication that the bias increases with the redshift, as was also observed with the data from HSC DR3. The third-party dataset therefore shows the same observation as the dataset of HSC DR3 galaxies and in fact data from all other premier sky surveys.

\section{Possible link to Zwicky's ``tired-light'' model}
\label{tired_light}

The early Universe as imaged by JWST is different than the early Universe predicted by the standard model. Among other explanations, that tension was proposed to have a link to Fritz Zwicky's ``tired-light'' (TL) theory \citep{zwicky1929redshift,yourgrau1971tired,vigier1977cosmological,shao2013energy,kragh2017universe,shao2018tired,sato2019tired,laviolette2021expanding,lovyagin2022cosmological,lopez2023history,gupta2023jwst,gupta2024dark}.

According to TL, photons lose their energy along their traveling path through the Universe. That can lead to differences in the redshift as observed from Earth, and therefore different galaxies that are more distant from Earth can have different redshift than galaxies that are closer to Earth. If that theory is correct, it can explain the early mature galaxies observed by JWST, as their true age is not the same age that their redshift indicates. In its extreme form, TL can argue that the Big Bang is merely an artifact created by TL, and the Universe is in fact in steady state. One of the downsides of the model is that there is no empirical observation that can show that photons indeed lose their energy as they travel through the Cosmos.

An empirical experiment that shows a redshift bias that grows as the redshift gets higher is difficult to perform at cosmological-scale distance, since the only information that can be obtained is the redshift as observed from Earth. While the distances of the galaxies can be estimated through other candles such as Ia supernovae, it is difficult to prove whether the redshift of a galaxy is higher because the galaxy indeed moves faster away from Earth in accordance to the Big Bang theory, because of TL, or a combination of the two. Ia supernovae are also limited in the distance range compared to the redshift, and therefore a redshift bias in the very high redshifts will be difficult to profile using Ia supernovae.

But when using the rotational velocity of the Earth within the Milky Way galaxy, a small redshift bias is expected due to the rotational velocity of the Earth. That bias is expected to affect galaxies that rotate in the same direction relative to the Milky Way differently than it affect galaxies that rotate in the opposite direction relative to the Milky Way. That bias is expected to be small, but according to several datasets as shown here it is statistically significant. More importantly, the bias grows with the redshift, suggesting that it is not the velocity alone that leads to the bias. The rotational velocity of the Earth within the Milky Way galaxy is obviously a constant that does not change, and the radial velocity of galaxies that rotate in the same direction relative to the Milky Way is expected to be, on average, the same as the radial velocity of galaxies that rotate in the opposite direction relative to the Milky Way. The idea that the redshift can change with the distance, and not necessarily the velocity, might be an indication supporting Fritz Zwicky's century-old theory.    

That observation can also be related to the far higher number of galaxies that rotate in the opposite direction relative to the Milky Way as imaged by JWST \citep{shamir2024galaxy}. The JWST deep field images show a 140\% more galaxies that rotate in the opposite direction relative to the Milky Way \citep{shamir2024galaxy}. A possible explanation is that galaxies that rotate in the opposite direction relative to the Milky Way are brighter due to the Doppler shift effect. Due to TL, the brightness difference gets more significant when the galaxies are more distant from Earth, and therefore much more galaxies that rotate in the opposite direction relative to the Milky Way are observed. Earlier observation made before JWST saw first light also showed such asymmetry that grows with the redshift \citep{shamir2020patterns,shamir2022large}, although at the lower redshift ranged the asymmetry was far milder than the asymmetry observed in the early Universe image by JWST.

\section{Conclusion}
\label{conclusion}

Unexpected observations such as the $H_0$ tension and galaxies that according to the current theories are expected to be older than what traditional galaxy formation models predicted are challenging the standard cosmological model. If the cosmological model is complete and fully accurate, the distance measurements, and primarily the redshift, are biased. If the redshift is fully accurate then the standard cosmological model and basic theories regarding galaxy formation and the history of the Universe are incomplete. In any case, the redshift as used currently and the existing basic cosmological theories cannot co-exist without modifications.

This paper presents empirical observations that show that the redshift model might be biased, and the bias might be driven by the rotational velocity of the Milky Way relative to the rotational velocity of the observed galaxies. The observed bias is consistent across different telescopes, different annotation methods, and shows very similar bias in both ends of the Galactic pole. It is also consistent in catalogs that were collected for other purposes by different research teams.  

The empirical observations described in this paper are provided with the data to ensure that the results can be reproduced. It had been shown that the vast majority of the scientific results cannot be reproduced \citep{stodden2018empirical}, introducing the challenge known as the ``reproducibility crisis'' in science \citep{baker2016reproducibility,miyakawa2020no,sayre2018reproducibility,ball2023}. The ability to access the data and reproduce the results allows to advance science in a transparent manner, and avoid errors that might not be noticeable to a reader unless they have access to the data. 

In the current astrophysics and cosmology practices, the redshift is used in most cases by ignoring the rotational velocity of the Milky Way, as the rotational velocity is far lower than the linear velocity and can therefore be considered negligible. But it should be reminded that the physics of galaxy rotation, and in particular the rotational velocity of galaxies, is still not fully understood \citep{opik1922estimate,babcock1939rotation,oort1940some,rubin1970rotation,rubin1978extended,rubin1980rotational,rubin1985rotation,sanders1990mass,rubin2000,mannheim2006alternatives,kroupa2012dark,kroupa2012failures,kroupa2015galaxies,arun2017dark,akerib2017,bertone2018new,aprile2018,skordis2019gravitational,sivaram2020mond,hofmeister2020debate,byrd2021spiral,gomel2021effects,haslbauer2022has}. . Theories such as dark matter \citep{rubin1983rotation} or MOND \citep{milgrom1983modification} have been proposed to explain the anomaly of the rotational velocity of galaxies, but several decades of research still have not let to a proven explanation to the provocative nature of the rotational velocity of galaxies.

It is difficult to identify an immediate explanation to the link between the rotational velocity and the redshift as observed from Earth. A possible explanation is the tired-light theory. But as mentioned above, the physics of galaxy rotation in general is difficult to explain without making unproven assumptions. Since the redshift is the most common distance indicator in cosmological scales, a bias in the redshift can impact a large number of other studies that make use of the redshift.  

Because the bias tends to becomes larger when the redshift gets higher, it is possible that such bias can explain anomalies such as galaxies that according to the existing theories are expected to be older than what traditional galaxy formation models predicted. The experiments described here were based on relatively low redshift ranges, and therefore it is still unclear whether higher redshifts will have significant redshift bias. Studying the bias in higher redshifts will require to use a large number of galaxies with redshift imaged by space-based instruments such as JWST at around the galactic pole.

Because $H_0$ is determined by using the redshift, a redshift bias can also explain the observed $H_0$ tension. For instance, when using the {\it SH0ES} catalog \citep{refId0} of Ia supernovae by just selecting the galaxies that rotate in the same direction as the Milky Way, $H_0$ drops from $\sim$73.7 to $\sim$69.05 $km/s Mpc^{-1}$ \citep{mcadam2023asymmetry}, which is within statistical error from the $H_0$ as observed by the CMB. When using just {\it SH0ES} galaxies that rotate in the opposite direction relative to the Milky Way, the $H_0$ increases to $\sim$74.2 $km/s Mpc^{-1}$ \citep{mcadam2023asymmetry}. Although {\it SH0ES} contains a relatively small number of Ia supernovae with their host galaxies, it suggests that the redshift as a distance indicator might depend on the rotational velocity relative to the rotational velocity of the Milky Way. This observation is also aligned with the contention the $H_0$ tension might require new physics that applies to the entire Universe, rather than certain changes in the physics of the early Universe \citep{vagnozzi2023seven}. Because the $H_0$ is determined by using the redshift, redshift bias can also be related to the observed $H_0$ anisotropy \citep{javanmardi2015probing,krishnan2022hints,cowell2022potential,cowell2022potential2,aluri2023observable}, which is another puzzling observation that does not have an immediate explanation.

It is also possible that the redshift difference is not a bias, and galaxies that rotate in the opposite direction relative to the Milky Way are indeed closer to Earth compared to galaxies that rotate in the same direction relative to the Milky Way. In that case the alignment with both ends of the Galactic pole is merely a coincidence. Such large-scale alignment is far larger than any known cluster, super-cluster, or filament in the cosmic web. That might be in agreement with numerous other observations that suggest that the cosmological principle is violated \citep{aluri2023observable}.

Although alignment in galaxy spin directions is expected \citep{d2022intrinsic,kraljic2020and}, it does not expect to form a cosmological-scale axis. If such axis indeed exists and it is not driven by the impact of the rotational velocity on the redshift measurements, it can be linked with theories such as dipole cosmology \citep{ebrahimian2023towards,krishnan2022tilt,allahyari2023big,krishnan2023dipole,krishnan2023copernican}, or rotating universe \citep{godel1949example,ozsvath1962finite,godel2000rotating,chechin2016rotation,camp2021}. Theories that assume a universe rotating around a cosmological-scale axis include Black Hole Cosmology \citep{pathria1972universe,stuckey1994observable,easson2001universe,poplawski2010radial,christillin2014machian,dymnikova2019universes,chakrabarty2020toy,poplawski2021nonsingular,gaztanaga2022black,gaztanaga2022black2}  and ellipsoidal universe \citep{campanelli2006ellipsoidal,campanelli2007cosmic,gruppuso2007complete,campanelli2011cosmic,cea2014ellipsoidal}.

Tensions between the expected age of some galaxies and the age of the Universe, as well as other cosmological-scale anisotropies and observations such as the $H_o$ tension, challenge our understanding of the Universe. It is clear that the current theories cannot co-exist with the reshift model as it is used currently, and therefore if the current theories are complete, it means that the redshift as a distance indicator is incomplete. This paper showed consistent evidence that the redshift depends on the rotational velocity of the Milky Way relative to the observed objects. The bias is small, but if it increases in the redshift ranges of the JWST deep fields it can potentially explain the existence of mature galaxies in the early Universe.

\section*{Data availability}

All datasets used in this paper are publicly available. HSC DR3 galaxy data are available at \url{https://people.cs.ksu.edu/~lshamir/data/zdifference_hsc/}. Annotated SDSS and DESI galaxies from the Northern and Southern Galactic poles are available at \url{https://people.cs.ksu.edu/~lshamir/data/zdifference/}. The data used by Michael Longo can be accessed at \url{https://ars.els-cdn.com/content/image/1-s2.0-S0370269311003947-mmc1.txt}. SDSS galaxies annotated by {\it SPARCFIRE} are available at \url{https://people.cs.ksu.edu/~lshamir/data/sparcfire/}. {\it Galaxy Zoo} data are available through SDSS CAS at \url{http://casjobs.sdss.org/CasJobs/default.aspx}.

\section*{Acknowledgments}

I would like to thank the three anonymous reviewers and the associate editor for the helpful comments. This study was supported in part by NSF grant 2148878.

\bibliographystyle{apalike}
\bibliography{main}

\begin{thebibliography}{}

\bibitem[Akerib et~al., 2017]{akerib2017}
Akerib, D.~S., Alsum, S., Ara\'ujo, H.~M., Bai, X., Bailey, A.~J., Balajthy,
  J., Beltrame, P., Bernard, E.~P., Bernstein, A., Biesiadzinski, T.~P.,
  Boulton, E.~M., Bramante, R., Br\'as, P., Byram, D., Cahn, S.~B.,
  Carmona-Benitez, M.~C., Chan, C., Chiller, A.~A., Chiller, C., Currie, A.,
  Cutter, J.~E., Davison, T. J.~R., Dobi, A., Dobson, J. E.~Y., Druszkiewicz,
  E., Edwards, B.~N., Faham, C.~H., Fiorucci, S., Gaitskell, R.~J., Gehman,
  V.~M., Ghag, C., Gibson, K.~R., Gilchriese, M. G.~D., Hall, C.~R., Hanhardt,
  M., Haselschwardt, S.~J., Hertel, S.~A., Hogan, D.~P., Horn, M., Huang,
  D.~Q., Ignarra, C.~M., Ihm, M., Jacobsen, R.~G., Ji, W., Kamdin, K., Kazkaz,
  K., Khaitan, D., Knoche, R., Larsen, N.~A., Lee, C., Lenardo, B.~G., Lesko,
  K.~T., Lindote, A., Lopes, M.~I., Manalaysay, A., Mannino, R.~L., Marzioni,
  M.~F., McKinsey, D.~N., Mei, D.-M., Mock, J., Moongweluwan, M., Morad, J.~A.,
  Murphy, A. S.~J., Nehrkorn, C., Nelson, H.~N., Neves, F., O'Sullivan, K.,
  Oliver-Mallory, K.~C., Palladino, K.~J., Pease, E.~K., Phelps, P., Reichhart,
  L., Rhyne, C., Shaw, S., Shutt, T.~A., Silva, C., Solmaz, M., Solovov, V.~N.,
  Sorensen, P., Stephenson, S., Sumner, T.~J., Szydagis, M., Taylor, D.~J.,
  Taylor, W.~C., Tennyson, B.~P., Terman, P.~A., Tiedt, D.~R., To, W.~H.,
  Tripathi, M., Tvrznikova, L., Uvarov, S., Verbus, J.~R., Webb, R.~C., White,
  J.~T., Whitis, T.~J., Witherell, M.~S., Wolfs, F. L.~H., Xu, J., Yazdani, K.,
  Young, S.~K., and Zhang, C. (2017).
\newblock Results from a search for dark matter in the complete lux exposure.
\newblock {\em Phys. Rev. Lett.}, 118:021303.

\bibitem[Allahyari et~al., 2023]{allahyari2023big}
Allahyari, A., Ebrahimian, E., Mondol, R., and Sheikh-Jabbari, M. (2023).
\newblock Big bang in dipole cosmology.
\newblock {\em arXiv:2307.15791}.

\bibitem[Aluri et~al., 2023]{aluri2023observable}
Aluri, P.~K., Cea, P., Chingangbam, P., Chu, M.-C., Clowes, R.~G.,
  Hutsemékers, D., Kochappan, J.~P., Krasiński, A., Lopez, A.~M., Liu, L.,
  Martens, N. C.~M., Martins, C. J. A.~P., Migkas, K., Colgáin, E.~., Pranav,
  P., Shamir, L., Singal, A.~K., Sheikh-Jabbari, M.~M., Wagner, J., Wang,
  S.-J., Wiltshire, D.~L., Yeung, S., Yin, L., and Zhao, W. (2023).
\newblock Is the observable universe consistent with the cosmological
  principle?
\newblock {\em Classical and Quantum Gravity}, 40(9):094001.

\bibitem[Aprile et~al., 2018]{aprile2018}
Aprile, E., Aalbers, J., Agostini, F., Alfonsi, M., Althueser, L., Amaro,
  F.~D., Anthony, M., Arneodo, F., Baudis, L., Bauermeister, B.,
  Benabderrahmane, M.~L., Berger, T., Breur, P.~A., Brown, A., Brown, A.,
  Brown, E., Bruenner, S., Bruno, G., Budnik, R., Capelli, C., Cardoso, J.
  M.~R., Cichon, D., Coderre, D., Colijn, A.~P., Conrad, J., Cussonneau, J.~P.,
  Decowski, M.~P., de~Perio, P., Di~Gangi, P., Di~Giovanni, A., Diglio, S.,
  Elykov, A., Eurin, G., Fei, J., Ferella, A.~D., Fieguth, A., Fulgione, W.,
  Gallo~Rosso, A., Galloway, M., Gao, F., Garbini, M., Geis, C., Grandi, L.,
  Greene, Z., Qiu, H., Hasterok, C., Hogenbirk, E., Howlett, J., Itay, R.,
  Joerg, F., Kaminsky, B., Kazama, S., Kish, A., Koltman, G., Landsman, H.,
  Lang, R.~F., Levinson, L., Lin, Q., Lindemann, S., Lindner, M., Lombardi, F.,
  Lopes, J. A.~M., Mahlstedt, J., Manfredini, A., Marrod\'an~Undagoitia, T.,
  Masbou, J., Masson, D., Messina, M., Micheneau, K., Miller, K., Molinario,
  A., Mor\aa{}, K., Murra, M., Naganoma, J., Ni, K., Oberlack, U., Pelssers,
  B., Piastra, F., Pienaar, J., Pizzella, V., Plante, G., Podviianiuk, R.,
  Priel, N., Ram\'{\i}rez~Garc\'{\i}a, D., Rauch, L., Reichard, S., Reuter, C.,
  Riedel, B., Rizzo, A., Rocchetti, A., Rupp, N., dos Santos, J. M.~F.,
  Sartorelli, G., Scheibelhut, M., Schindler, S., Schreiner, J., Schulte, D.,
  Schumann, M., Scotto~Lavina, L., Selvi, M., Shagin, P., Shockley, E., Silva,
  M., Simgen, H., Thers, D., Toschi, F., Trinchero, G., Tunnell, C., Upole, N.,
  Vargas, M., Wack, O., Wang, H., Wang, Z., Wei, Y., Weinheimer, C., Wittweg,
  C., Wulf, J., Ye, J., Zhang, Y., and Zhu, T. (2018).
\newblock Dark matter search results from a one ton-year exposure of xenon1t.
\newblock {\em Phys. Rev. Lett.}, 121:111302.

\bibitem[Arun et~al., 2017]{arun2017dark}
Arun, K., Gudennavar, S., and Sivaram, C. (2017).
\newblock Dark matter, dark energy, and alternate models: A review.
\newblock {\em Advances in Space Research}, 60(1):166--186.

\bibitem[Babcock, 1939]{babcock1939rotation}
Babcock, H.~W. (1939).
\newblock The rotation of the andromeda nebula.
\newblock {\em Lick Observatory Bulletin}, 19:41--51.

\bibitem[Baker, 2016]{baker2016reproducibility}
Baker, M. (2016).
\newblock Reproducibility crisis.
\newblock {\em nature}, 533(26):353--66.

\bibitem[Ball, 2023]{ball2023}
Ball, P. (2023).
\newblock Is {AI} leading to a reproducibility crisis in science?
\newblock {\em Nature}, 624:22--25.

\bibitem[Bertone and Tait, 2018]{bertone2018new}
Bertone, G. and Tait, T.~M. (2018).
\newblock A new era in the search for dark matter.
\newblock {\em Nature}, 562(7725):51--56.

\bibitem[Blake, 2021]{blake2021relativistic}
Blake, B.~C. (2021).
\newblock Relativistic beaming of gravity and the missing mass problem.
\newblock {\em Bulletin of the American Physical Society}, page B17.00002.

\bibitem[Bolejko, 2018]{bolejko2018emerging}
Bolejko, K. (2018).
\newblock Emerging spatial curvature can resolve the tension between
  high-redshift cmb and low-redshift distance ladder measurements of the hubble
  constant.
\newblock {\em Physical Review D}, 97(10):103529.

\bibitem[Buta et~al., 2003]{buta2003ringed}
Buta, R.~J., Byrd, G.~G., and Freeman, T. (2003).
\newblock The ringed spiral galaxy ngc 4622. i. photometry, kinematics, and the
  case for two strong leading outer spiral arms.
\newblock {\em Astronomical Journal}, 125(2):634.

\bibitem[Byrd and Howard, 2021]{byrd2021spiral}
Byrd, G. and Howard, S. (2021).
\newblock Spiral galaxies when disks dominate their halos (using arm pitches
  and rotation curves).
\newblock {\em Journal of the Washington Academy of Sciences}, 107(1):1.

\bibitem[Camarena and Marra, 2020]{camarena2020local}
Camarena, D. and Marra, V. (2020).
\newblock Local determination of the hubble constant and the deceleration
  parameter.
\newblock {\em Physical Review Research}, 2(1):013028.

\bibitem[Campanelli, 2021]{camp2021}
Campanelli, L. (2021).
\newblock A conjecture on the neutrality of matter.
\newblock {\em Foundations of Physics}, 51:56.

\bibitem[Campanelli et~al., 2011]{campanelli2011cosmic}
Campanelli, L., Cea, P., Fogli, G., and Tedesco, L. (2011).
\newblock Cosmic parallax in ellipsoidal universe.
\newblock {\em Modern Physics Letters A}, 26(16):1169--1181.

\bibitem[Campanelli et~al., 2006]{campanelli2006ellipsoidal}
Campanelli, L., Cea, P., and Tedesco, L. (2006).
\newblock Ellipsoidal universe can solve the cosmic microwave background
  quadrupole problem.
\newblock {\em PRL}, 97(13):131302.

\bibitem[Campanelli et~al., 2007]{campanelli2007cosmic}
Campanelli, L., Cea, P., and Tedesco, L. (2007).
\newblock Cosmic microwave background quadrupole and ellipsoidal universe.
\newblock {\em Physical Review D}, 76(6):063007.

\bibitem[Capozziello and De~Laurentis, 2012]{capozziello2012dark}
Capozziello, S. and De~Laurentis, M. (2012).
\newblock The dark matter problem from f (r) gravity viewpoint.
\newblock {\em Annalen der Physik}, 524(9-10):545--578.

\bibitem[Cea, 2014]{cea2014ellipsoidal}
Cea, P. (2014).
\newblock The ellipsoidal universe in the planck satellite era.
\newblock {\em Monthly Notices of the Royal Astronomical Society},
  441(2):1646--1661.

\bibitem[Chadwick et~al., 2013]{chadwick2013gravitational}
Chadwick, E.~A., Hodgkinson, T.~F., and McDonald, G.~S. (2013).
\newblock Gravitational theoretical development supporting mond.
\newblock {\em Physical Review D}, 88(2):024036.

\bibitem[Chakrabarty et~al., 2020]{chakrabarty2020toy}
Chakrabarty, H., Abdujabbarov, A., Malafarina, D., and Bambi, C. (2020).
\newblock A toy model for a baby universe inside a black hole.
\newblock {\em European Physical Journal C}, 80(1909.07129):1--10.

\bibitem[Chechin, 2016]{chechin2016rotation}
Chechin, L. (2016).
\newblock Rotation of the universe at different cosmological epochs.
\newblock {\em Astronomy Reports}, 60(6):535--541.

\bibitem[Christillin, 2014]{christillin2014machian}
Christillin, P. (2014).
\newblock The machian origin of linear inertial forces from our gravitationally
  radiating black hole universe.
\newblock {\em The European Physical Journal Plus}, 129(8):1--3.

\bibitem[Cowell et~al., 2022]{cowell2022potential}
Cowell, J.~A., Dhawan, S., and Macpherson, H.~J. (2022).
\newblock Potential signature of a quadrupolar hubble expansion in pantheon+
  supernovae.
\newblock {\em arXiv:2212.13569}.

\bibitem[Cowley et~al., 2018]{cowley2018predictions}
Cowley, W.~I., Baugh, C.~M., Cole, S., Frenk, C.~S., and Lacey, C.~G. (2018).
\newblock Predictions for deep galaxy surveys with {JWST} from $\lambda$cdm.
\newblock {\em Monthly Notices of the Royal Astronomical Society},
  474(2):2352--2372.

\bibitem[Crawford, 1999]{crawford1999curvature}
Crawford, D.~F. (1999).
\newblock Curvature pressure in a cosmology with a tired-light redshift.
\newblock {\em Australian Journal of Physics}, 52(4):753--777.

\bibitem[Curtis-Lake et~al., 2023]{curtis2023spectroscopic}
Curtis-Lake, E., Carniani, S., Cameron, A., Charlot, S., Jakobsen, P.,
  Maiolino, R., Bunker, A., Witstok, J., Smit, R., Chevallard, J., et~al.
  (2023).
\newblock Spectroscopic confirmation of four metal-poor galaxies at z=
  10.3--13.2.
\newblock {\em Nature Astronomy}, 7(5):622--632.

\bibitem[d’Assignies D et~al., 2022]{d2022intrinsic}
d’Assignies D, W., Chisari, N.~E., Hamaus, N., and Singh, S. (2022).
\newblock Intrinsic alignments of galaxies around cosmic voids.
\newblock {\em Monthly Notices of the Royal Astronomical Society},
  509(2):1985--1994.

\bibitem[Davis and Hayes, 2021]{sparcfire_ascl}
Davis, D. and Hayes, W. (2021).
\newblock Sparcfire: Spiral arc finder and reporter.
\newblock {\em ASCL}, page ascl:2107.010.

\bibitem[Davis and Hayes, 2014]{Davis_2014}
Davis, D.~R. and Hayes, W.~B. (2014).
\newblock {SpArcFiRe}: {Scalable} {Automated} {Detection} {of} {Spiral}
  {Galaxy} {Arm} {Segments}.
\newblock {\em Astrophysical Journal}, 790(2):87.

\bibitem[Davis et~al., 2019]{davis2019can}
Davis, T.~M., Hinton, S.~R., Howlett, C., and Calcino, J. (2019).
\newblock Can redshift errors bias measurements of the hubble constant?
\newblock {\em Monthly Notices of the Royal Astronomical Society},
  490(2):2948--2957.

\bibitem[Di~Valentino et~al., 2021]{di2021realm}
Di~Valentino, E., Mena, O., Pan, S., Visinelli, L., Yang, W., Melchiorri, A.,
  Mota, D.~F., Riess, A.~G., and Silk, J. (2021).
\newblock In the realm of the hubble tension—a review of solutions.
\newblock {\em Classical and Quantum Gravity}, 38(15):153001.

\bibitem[Dymnikova, 2019]{dymnikova2019universes}
Dymnikova, I. (2019).
\newblock Universes inside a black hole with the de sitter interior.
\newblock {\em Universe}, 5(5):111.

\bibitem[Easson and Brandenberger, 2001]{easson2001universe}
Easson, D.~A. and Brandenberger, R.~H. (2001).
\newblock Universe generation from black hole interiors.
\newblock {\em Journal of High Energy Physics}, 2001(06):024.

\bibitem[Ebrahimian et~al., 2023]{ebrahimian2023towards}
Ebrahimian, E., Krishnan, C., Mondol, R., and Sheikh-Jabbari, M. (2023).
\newblock Towards a realistic dipole cosmology: The dipole $\backslash$lambda
  cdm model.
\newblock {\em arXiv:2305.16177}.

\bibitem[El-Neaj et~al., 2020]{el2020aedge}
El-Neaj, Y.~A., Alpigiani, C., Amairi-Pyka, S., Ara{\'u}jo, H., Bala{\v{z}},
  A., Bassi, A., Bathe-Peters, L., Battelier, B., Beli{\'c}, A., Bentine, E.,
  et~al. (2020).
\newblock Aedge: atomic experiment for dark matter and gravity exploration in
  space.
\newblock {\em EPJ Quantum Technology}, 7(1):1--27.

\bibitem[Farnes, 2018]{farnes2018unifying}
Farnes, J.~S. (2018).
\newblock A unifying theory of dark energy and dark matter: Negative masses and
  matter creation within a modified $\lambda$cdm framework.
\newblock {\em Astronomy \& Astrophysics}, 620:A92.

\bibitem[Gaztanaga, 2022a]{gaztanaga2022black}
Gaztanaga, E. (2022a).
\newblock The black hole universe, part i.
\newblock {\em Symmetry}, 14(9):1849.

\bibitem[Gaztanaga, 2022b]{gaztanaga2022black2}
Gaztanaga, E. (2022b).
\newblock The black hole universe, part ii.
\newblock {\em Symmetry}, 14(10):1984.

\bibitem[Glazebrook et~al., 2024]{glazebrook2024massive}
Glazebrook, K., Nanayakkara, T., Schreiber, C., Lagos, C., Kawinwanichakij, L.,
  Jacobs, C., Chittenden, H., Brammer, G., Kacprzak, G.~G., Labbe, I., et~al.
  (2024).
\newblock A massive galaxy that formed its stars at z\~{} 11.
\newblock {\em Nature}, pages 1--3.

\bibitem[G{\"o}del, 1949]{godel1949example}
G{\"o}del, K. (1949).
\newblock An example of a new type of cosmological solutions of einstein's
  field equations of gravitation.
\newblock {\em Reviews of Modern Physics}, 21(3):447.

\bibitem[G{\"o}del, 2000]{godel2000rotating}
G{\"o}del, K. (2000).
\newblock Rotating universes in general relativity theory.
\newblock {\em General Relativity and Gravitation}, 32(7):1419--1427.

\bibitem[Gomel and Zimmerman, 2021]{gomel2021effects}
Gomel, R. and Zimmerman, T. (2021).
\newblock The effects of inertial forces on the dynamics of disk galaxies.
\newblock {\em Galaxies}, 9(2):34.

\bibitem[Gruppuso, 2007]{gruppuso2007complete}
Gruppuso, A. (2007).
\newblock Complete statistical analysis for the quadrupole amplitude in an
  ellipsoidal universe.
\newblock {\em Physical Review D}, 76(8):083010.

\bibitem[Gupta, 2023]{gupta2023jwst}
Gupta, R. (2023).
\newblock {JWST} early universe observations and $\lambda$cdm cosmology.
\newblock {\em Monthly Notices of the Royal Astronomical Society}, page
  stad2032.

\bibitem[Gupta, 2024]{gupta2024dark}
Gupta, R.~P. (2024).
\newblock On dark matter and dark energy in ccc+ tl cosmology.
\newblock {\em Universe}, 10(6):266.

\bibitem[Haslbauer et~al., 2022a]{haslbauer2022high}
Haslbauer, M., Banik, I., Kroupa, P., Wittenburg, N., and Javanmardi, B.
  (2022a).
\newblock The high fraction of thin disk galaxies continues to challenge
  $\lambda$cdm cosmology.
\newblock {\em Astrophysical Journal}, 925(2):183.

\bibitem[Haslbauer et~al., 2022b]{haslbauer2022has}
Haslbauer, M., Kroupa, P., Zonoozi, A.~H., and Haghi, H. (2022b).
\newblock Has {JWST} already falsified dark-matter-driven galaxy formation?
\newblock {\em Astrophysical Journal Letters}, 939(2):L31.

\bibitem[Hayes et~al., 2017]{hayes2017nature}
Hayes, W.~B., Davis, D., and Silva, P. (2017).
\newblock On the nature and correction of the spurious s-wise spiral galaxy
  winding bias in galaxy zoo 1.
\newblock {\em Monthyl Notices of the Royal Astronomical Society},
  466(4):3928--3936.

\bibitem[Hofmeister and Criss, 2020]{hofmeister2020debate}
Hofmeister, A.~M. and Criss, R.~E. (2020).
\newblock Debate on the physics of galactic rotation and the existence of dark
  matter.

\bibitem[Javanmardi et~al., 2015]{javanmardi2015probing}
Javanmardi, B., Porciani, C., Kroupa, P., and Pflam-Altenburg, J. (2015).
\newblock Probing the isotropy of cosmic acceleration traced by type ia
  supernovae.
\newblock {\em Astrophysical Journal}, 810(1):47.

\bibitem[Khetan et~al., 2021]{refId0}
Khetan, N., Izzo, L., Branchesi, M., Wojtak, R., Cantiello, M., Murugeshan, C.,
  Agnello, A., Cappellaro, E., Della~Valle, M., Gall, C., Hjorth, J., Benetti,
  S., Brocato, E., Burke, J., Hiramatsu, D., Howell, D.~A., Tomasella, L., and
  Valenti, S. (2021).
\newblock A new measurement of the hubble constant using type ia supernovae
  calibrated with surface brightness fluctuations.
\newblock {\em Astronomy \& Astrophysics}, 647:A72.

\bibitem[Kragh, 2017]{kragh2017universe}
Kragh, H. (2017).
\newblock Is the universe expanding? fritz zwicky and early tired-light
  hypotheses.
\newblock {\em Journal of Astronomical History and Heritage}, 20(1):2--12.

\bibitem[Kraljic et~al., 2020]{kraljic2020and}
Kraljic, K., Dav{\'e}, R., and Pichon, C. (2020).
\newblock And yet it flips: connecting galactic spin and the cosmic web.
\newblock {\em Monthly Notices of the Royal Astronomical Society},
  493(1):362--381.

\bibitem[Krishnan et~al., 2022a]{krishnan2022hints}
Krishnan, C., Mohayaee, R., Colg{\'a}in, E.~{\'O}., Sheikh-Jabbari, M., and
  Yin, L. (2022a).
\newblock Hints of flrw breakdown from supernovae.
\newblock {\em Physical Review D}, 105(6):063514.

\bibitem[Krishnan et~al., 2023a]{krishnan2023copernican}
Krishnan, C., Mondol, R., and Jabbari, M.~S. (2023a).
\newblock Copernican paradigm beyond flrw.

\bibitem[Krishnan et~al., 2022b]{krishnan2022tilt}
Krishnan, C., Mondol, R., and Sheikh-Jabbari, M. (2022b).
\newblock A tilt instability in the cosmological principle.
\newblock {\em arXiv:2211.08093}.

\bibitem[Krishnan et~al., 2023b]{krishnan2023dipole}
Krishnan, C., Mondol, R., and Sheikh-Jabbari, M. (2023b).
\newblock Dipole cosmology: the copernican paradigm beyond flrw.
\newblock {\em Journal of Cosmology and Astroparticle Physics}, 2023(07):020.

\bibitem[Kroupa, 2012]{kroupa2012dark}
Kroupa, P. (2012).
\newblock The dark matter crisis: falsification of the current standard model
  of cosmology.
\newblock {\em Publications of the Astronomical Society of Australia},
  29(4):395--433.

\bibitem[Kroupa, 2015]{kroupa2015galaxies}
Kroupa, P. (2015).
\newblock Galaxies as simple dynamical systems: observational data disfavor
  dark matter and stochastic star formation.
\newblock {\em Canadian Journal of Physics}, 93(2):169--202.

\bibitem[Kroupa et~al., 2012]{kroupa2012failures}
Kroupa, P., Pawlowski, M., and Milgrom, M. (2012).
\newblock The failures of the standard model of cosmology require a new
  paradigm.
\newblock {\em International Journal of Modern Physics D}, 21(14):1230003.

\bibitem[Larin, 2022]{larin2022towards}
Larin, S.~A. (2022).
\newblock Towards the explanation of flatness of galaxies rotation curves.
\newblock {\em Universe}, 8(12):632.

\bibitem[Lavery et~al., 2014]{lavery2014observation}
Lavery, M.~P., Barnett, S.~M., Speirits, F.~C., and Padgett, M.~J. (2014).
\newblock Observation of the rotational {D}oppler shift of a white-light,
  orbital-angular-momentum-carrying beam backscattered from a rotating body.
\newblock {\em Optica}, 1(1):1--4.

\bibitem[LaViolette, 2021]{laviolette2021expanding}
LaViolette, P.~A. (2021).
\newblock Expanding or static universe: Emergence of a new paradigm.
\newblock {\em International Journal of Astronomy and Astrophysics},
  11(2):190--231.

\bibitem[Lee, 2023]{lee2023cosmological}
Lee, S. (2023).
\newblock The cosmological evolution condition of the {P}lanck constant in the
  varying speed of light models through adiabatic expansion.
\newblock {\em Physics of the Dark Universe}, page 101286.

\bibitem[Lintott et~al., 2008]{lintott2008galaxy}
Lintott, C.~J., Schawinski, K., Slosar, A., Land, K., Bamford, S., Thomas, D.,
  Raddick, M.~J., Nichol, R.~C., Szalay, A., Andreescu, D., et~al. (2008).
\newblock Galaxy zoo: morphologies derived from visual inspection of galaxies
  from the sloan digital sky survey.
\newblock {\em Monthly Notices of the Royal Astronomical Society},
  389(3):1179--1189.

\bibitem[Liu et~al., 2019]{liu2019experimental}
Liu, B., Chu, H., Giddens, H., Li, R., and Hao, Y. (2019).
\newblock Experimental observation of linear and rotational {D}oppler shifts
  from several designer surfaces.
\newblock {\em Scientific Reports}, 9(1):8971.

\bibitem[Longo, 2011]{longo2011detection}
Longo, M.~J. (2011).
\newblock Detection of a dipole in the handedness of spiral galaxies with
  redshifts z~ 0.04.
\newblock {\em Physics Letters B}, 699(4):224--229.

\bibitem[Lopez-Corredoira, 2023]{lopez2023history}
Lopez-Corredoira, M. (2023).
\newblock History and problems of the standard model in cosmology.
\newblock {\em arXiv:2307.10606}.

\bibitem[Lovyagin et~al., 2022]{lovyagin2022cosmological}
Lovyagin, N., Raikov, A., Yershov, V., and Lovyagin, Y. (2022).
\newblock Cosmological model tests with jwst.
\newblock {\em Galaxies}, 10(6):108.

\bibitem[Mannheim, 2006]{mannheim2006alternatives}
Mannheim, P.~D. (2006).
\newblock Alternatives to dark matter and dark energy.
\newblock {\em Progress in Particle and Nuclear Physics}, 56(2):340--445.

\bibitem[Marrucci, 2013]{marrucci2013spinning}
Marrucci, L. (2013).
\newblock Spinning the {D}oppler effect.
\newblock {\em Science}, 341(6145):464--465.

\bibitem[McAdam and Shamir, 2023]{mcadam2023asymmetry}
McAdam, D. and Shamir, L. (2023).
\newblock Asymmetry between galaxy apparent magnitudes shows a possible tension
  between physical properties of galaxies and their rotational velocity.
\newblock {\em Symmetry}, 15(6):1190.

\bibitem[McConville and Colgain, 2023]{cowell2022potential2}
McConville, R. and Colgain, E. (2023).
\newblock Anisotropic distance ladder in pantheon+ supernovae.
\newblock {\em arXiv:2304.02718}.

\bibitem[Milgrom, 1983]{milgrom1983modification}
Milgrom, M. (1983).
\newblock A modification of the newtonian dynamics as a possible alternative to
  the hidden mass hypothesis.
\newblock {\em Astrophysical Journal}, 270:365--370.

\bibitem[Miyakawa, 2020]{miyakawa2020no}
Miyakawa, T. (2020).
\newblock No raw data, no science: another possible source of the
  reproducibility crisis.

\bibitem[M{\"o}rtsell and Dhawan, 2018]{mortsell2018does}
M{\"o}rtsell, E. and Dhawan, S. (2018).
\newblock Does the hubble constant tension call for new physics?
\newblock {\em Journal of Cosmology and Astroparticle Physics}, 2018(09):025.

\bibitem[Nagao, 2020]{nagao2020galactic}
Nagao, S. (2020).
\newblock Galactic evolution showing a constant circulating speed of stars in a
  galactic disc without requiring dark matter.
\newblock {\em Reports in Advances of Physical Sciences}, 4(02):2050004.

\bibitem[Neeleman et~al., 2020]{neeleman2020cold}
Neeleman, M., Prochaska, J.~X., Kanekar, N., and Rafelski, M. (2020).
\newblock A cold, massive, rotating disk galaxy 1.5 billion years after the big
  bang.
\newblock {\em Nature}, 581(7808):269--272.

\bibitem[Oort, 1940]{oort1940some}
Oort, J.~H. (1940).
\newblock Some problems concerning the structure and dynamics of the galactic
  system and the elliptical nebulae ngc 3115 and 4494.
\newblock {\em Astrophysical Journal}, 91:273.

\bibitem[Opik, 1922]{opik1922estimate}
Opik, E. (1922).
\newblock An estimate of the distance of the andromeda nebula.
\newblock {\em Astrophysical Journal}, 55.

\bibitem[Ozsv{\'a}th and Sch{\"u}cking, 1962]{ozsvath1962finite}
Ozsv{\'a}th, I. and Sch{\"u}cking, E. (1962).
\newblock Finite rotating universe.
\newblock {\em Nature}, 193(4821):1168--1169.

\bibitem[Pandey et~al., 2020]{pandey2020model}
Pandey, S., Raveri, M., and Jain, B. (2020).
\newblock Model independent comparison of supernova and strong lensing
  cosmography: Implications for the hubble constant tension.
\newblock {\em Physical Review D}, 102(2):023505.

\bibitem[Pathria, 1972]{pathria1972universe}
Pathria, R. (1972).
\newblock The universe as a black hole.
\newblock {\em Nature}, 240(5379):298--299.

\bibitem[Pletcher, 2023]{pletchermature}
Pletcher, A.~E. (2023).
\newblock Why mature galaxies seem to have filled the universe shortly after
  the big bang.
\newblock {\em Qeios}, page 10.32388/2X1GDL.2.

\bibitem[Pop{\l}awski, 2010]{poplawski2010radial}
Pop{\l}awski, N.~J. (2010).
\newblock Radial motion into an einstein--rosen bridge.
\newblock {\em Physics Letters B}, 687(2-3):110--113.

\bibitem[Pop{\l}awski, 2021]{poplawski2021nonsingular}
Pop{\l}awski, N.~J. (2021).
\newblock A nonsingular, anisotropic universe in a black hole with torsion and
  particle production.
\newblock {\em General Relativity and Gravitation}, 53(2):1--14.

\bibitem[Riess et~al., 2022]{riess2022comprehensive}
Riess, A.~G., Yuan, W., Macri, L.~M., Scolnic, D., Brout, D., Casertano, S.,
  Jones, D.~O., Murakami, Y., Anand, G.~S., Breuval, L., et~al. (2022).
\newblock A comprehensive measurement of the local value of the hubble constant
  with 1 km s$^{-1}$ {Mpc}$^{-1}$ uncertainty from the hubble space telescope
  and the sh0es team.
\newblock {\em Astrophysical Journal Letters}, 934(1):L7.

\bibitem[Rivera, 2020]{rivera2020alternative}
Rivera, P.~C. (2020).
\newblock An alternative model of rotation curve that explains anomalous
  orbital velocity, mass discrepancy and structure of some galaxies.
\newblock {\em American Journal of Astronomy and Astrophysics}, 7(4):73--79.

\bibitem[Rubin, 1983]{rubin1983rotation}
Rubin, V.~C. (1983).
\newblock The rotation of spiral galaxies.
\newblock {\em Science}, 220(4604):1339--1344.

\bibitem[Rubin et~al., 1985]{rubin1985rotation}
Rubin, V.~C., Burstein, D., Ford~Jr, W.~K., and Thonnard, N. (1985).
\newblock Rotation velocities of 16 sa galaxies and a comparison of sa, sb, and
  sc rotation properties.
\newblock {\em Astrophysical Journal}, 289:81--98.

\bibitem[Rubin and Ford~Jr, 1970]{rubin1970rotation}
Rubin, V.~C. and Ford~Jr, W.~K. (1970).
\newblock Rotation of the andromeda nebula from a spectroscopic survey of
  emission regions.
\newblock {\em Astrophysical Journal}, 159:379.

\bibitem[Rubin et~al., 1978]{rubin1978extended}
Rubin, V.~C., Ford~Jr, W.~K., and Thonnard, N. (1978).
\newblock Extended rotation curves of high-luminosity spiral galaxies.
  iv-systematic dynamical properties, sa through sc.
\newblock {\em Astrophysical Journal}, 225:L107--L111.

\bibitem[Rubin et~al., 1980]{rubin1980rotational}
Rubin, V.~C., Ford~Jr, W.~K., and Thonnard, N. (1980).
\newblock Rotational properties of 21 sc galaxies with a large range of
  luminosities and radii, from ngc 4605/r= 4kpc/to ugc 2885/r= 122 kpc.
\newblock {\em Astrophysical Journal}, 238:471--487.

\bibitem[Sanders, 1990]{sanders1990mass}
Sanders, R. (1990).
\newblock Mass discrepancies in galaxies: dark matter and alternatives.
\newblock {\em The Astronomy and Astrophysics Review}, 2(1):1--28.

\bibitem[Sato, 2019]{sato2019tired}
Sato, M. (2019).
\newblock Tired light: An alternative interpretation of the accelerating
  universe.
\newblock {\em Physics Essays}, 32(1):43--47.

\bibitem[Sayre and Riegelman, 2018]{sayre2018reproducibility}
Sayre, F. and Riegelman, A. (2018).
\newblock The reproducibility crisis and academic libraries.
\newblock {\em College \& Research Libraries}, 79(1):2.

\bibitem[Shamir, 2011a]{shamir2011ganalyzer}
Shamir, L. (2011a).
\newblock Ganalyzer: A tool for automatic galaxy image analysis.
\newblock {\em Astrophysical Journal}, 736(2):141.

\bibitem[Shamir, 2011b]{ganalyzer_ascl}
Shamir, L. (2011b).
\newblock Ganalyzer: A tool for automatic galaxy image analysis.
\newblock {\em ASCL}, page ascl:1105.011.

\bibitem[Shamir, 2020]{shamir2020patterns}
Shamir, L. (2020).
\newblock Patterns of galaxy spin directions in sdss and pan-starrs show parity
  violation and multipoles.
\newblock {\em Astrophysics \& Space Science}, 365:136.

\bibitem[Shamir, 2021]{shamir2021particles}
Shamir, L. (2021).
\newblock Analysis of the alignment of non-random patterns of spin directions
  in populations of spiral galaxies.
\newblock {\em Particles}, 4(1):11--28.

\bibitem[Shamir, 2022a]{shamir2022analysis}
Shamir, L. (2022a).
\newblock Analysis of spin directions of galaxies in the desi legacy survey.
\newblock {\em Monthly Notices of the Royal Astronomical Society},
  516(2):2281--2291.

\bibitem[Shamir, 2022b]{shamir2022large}
Shamir, L. (2022b).
\newblock Large-scale asymmetry in galaxy spin directions: analysis of galaxies
  with spectra in des, sdss, and desi legacy survey.
\newblock {\em Astronomical Notes}, 343(6-7):e20220010.

\bibitem[Shamir, 2024a]{shamir2024galaxy}
Shamir, L. (2024a).
\newblock Galaxy spin direction asymmetry in jwst deep fields.
\newblock {\em Publications of the Astronomical Society of Australia}, 41:e038.

\bibitem[Shamir, 2024b]{universe10030129}
Shamir, L. (2024b).
\newblock A simple direct empirical observation of systematic bias of the
  redshift as a distance indicator.
\newblock {\em Universe}, 10(3).

\bibitem[Shao, 2013]{shao2013energy}
Shao, M.-H. (2013).
\newblock The energy loss of photons and cosmological redshift.
\newblock {\em Physics Essays}, 26(2):183--190.

\bibitem[Shao et~al., 2018]{shao2018tired}
Shao, M.-H., Wang, N., and Gao, Z.-F. (2018).
\newblock Tired light denies the big bang.
\newblock In {\em Redefining Standard Model Cosmology}, pages 13--29.
  IntechOpen.

\bibitem[Sivaram et~al., 2020]{sivaram2020mond}
Sivaram, C., Arun, K., and Rebecca, L. (2020).
\newblock Mond, mong, morg as alternatives to dark matter and dark energy, and
  consequences for cosmic structures.
\newblock {\em Journal of Astrophysics and Astronomy}, 41(1):1--6.

\bibitem[Skordis and Z{\l}o{\'s}nik, 2019]{skordis2019gravitational}
Skordis, C. and Z{\l}o{\'s}nik, T. (2019).
\newblock Gravitational alternatives to dark matter with tensor mode speed
  equaling the speed of light.
\newblock {\em Physical Review D}, 100(10):104013.

\bibitem[Skordis and Z{\l}o{\'s}nik, 2021]{skordis2021new}
Skordis, C. and Z{\l}o{\'s}nik, T. (2021).
\newblock New relativistic theory for modified newtonian dynamics.
\newblock {\em Pattern Recognition Letters}, 127(16):161302.

\bibitem[Sofue and Rubin, 2001]{rubin2000}
Sofue, Y. and Rubin, V. (2001).
\newblock Rotation curves of spiral galaxies.
\newblock {\em Annual Review of Astronomy and Astrophysics}, 39:137--174.

\bibitem[Stodden et~al., 2018]{stodden2018empirical}
Stodden, V., Seiler, J., and Ma, Z. (2018).
\newblock An empirical analysis of journal policy effectiveness for
  computational reproducibility.
\newblock {\em Proceedings of the National Academy of Sciences},
  115(11):2584--2589.

\bibitem[Stuckey, 1994]{stuckey1994observable}
Stuckey, W. (1994).
\newblock The observable universe inside a black hole.
\newblock {\em American Journal of Physics}, 62(9):788--795.

\bibitem[Vagnozzi, 2023]{vagnozzi2023seven}
Vagnozzi, S. (2023).
\newblock Seven hints that early-time new physics alone is not sufficient to
  solve the hubble tension.
\newblock {\em Universe}, 9(9):393.

\bibitem[Vigier, 1977]{vigier1977cosmological}
Vigier, J.-P. (1977).
\newblock Cosmological implications of non-velocity redshifts—a tired-light
  mechanism.
\newblock In {\em Cosmology, History, and Theology}, pages 141--157. Springer.

\bibitem[Whitler et~al., 2023]{whitler2023ages}
Whitler, L., Endsley, R., Stark, D.~P., Topping, M., Chen, Z., and Charlot, S.
  (2023).
\newblock On the ages of bright galaxies~ 500 myr after the big bang: insights
  into star formation activity at z$>$15 with {JWST}.
\newblock {\em Monthly Notices of the Royal Astronomical Society},
  519(1):157--171.

\bibitem[Wu and Huterer, 2017]{wu2017sample}
Wu, H.-Y. and Huterer, D. (2017).
\newblock Sample variance in the local measurements of the hubble constant.
\newblock {\em Monthly Notices of the Royal Astronomical Society},
  471(4):4946--4955.

\bibitem[Yourgrau and Woodward, 1971]{yourgrau1971tired}
Yourgrau, W. and Woodward, J. (1971).
\newblock Tired light and the missing mass problem.
\newblock {\em Acta Physica Academiae Scientiarum Hungaricae}, 30(3):323--329.

\bibitem[Zwicky, 1929]{zwicky1929redshift}
Zwicky, F. (1929).
\newblock On the redshift of spectral lines through interstellar space.
\newblock {\em Proceedings of the National Academy of Sciences},
  15(10):773--779.

\end{thebibliography}

\end{document}